\pdfoutput=1
\documentclass[aps,prl,twocolumn,preprintnumbers,superscriptaddress,groupedaddress,nofootinbib]{revtex4-2}
\usepackage{bm}
\usepackage{amssymb}
\usepackage{amsmath,shuffle}

\allowdisplaybreaks

\def\eqref#1{(\ref{#1})}
\def\beq{\begin{equation}}
\def\eeq{\end{equation}}

\def\a{\alpha}
\def\b{{\beta}}
\def\g{{\gamma}}
\def\d{{\delta}}

\def\l{\lambda}

\def\t{{\theta}}

\def\p{{\partial}}
\def\({\left(}
\def\){\right)}

\def\ap{{\alpha'}}
\def\AFq(#1,#2,#3,#4,#5){A^{F^4}(#1,#2,#3,#4,#5)}

\begin{document}


\title{A relation between massive and massless string tree amplitudes}

\author{Sitender Pratap Kashyap$^a$, Carlos R. Mafra$^b$, Mritunjay Verma$^{b,c}$, and Luis
Alberto Ypanaqu\'e$^b$}
\affiliation{$^a$ Chennai Mathematical Institute, H1 SIPCOT IT Park, Kelambakkam, Tamil Nadu, India 603103}
\affiliation{$^b$ Mathematical Sciences and STAG Research Centre, University of Southampton, Highfield, Southampton, SO17 1BJ, UK}
\affiliation{$^c$ Indian Institute of Technology Indore, Khandwa Road, Simrol, Indore 453552, India}


\begin{abstract}
We uncover a relation
between the scattering amplitudes of massive strings and the $\ap$ expansion of
the massless string amplitude at tree level. More precisely, the $n$-point tree amplitude
of $n{-}1$ massless and one massive state is written as a linear combination of
$n{+}1$ massless string amplitudes at the $\ap^2$ order.
\end{abstract}

\maketitle

\section{\label{sec:intro}Introduction}

The spectrum of string theory contains an infinite tower of massive higher-spin states alongside
the massless excitations. These massive excitations are essential for the consistency of string
theory, such as perturbative unitarity. And yet, the calculation of scattering amplitudes
in the massive sector remains largely unexplored.

As a first line of attack, one may wish to accumulate data in the hopes of finding all-order
patterns. Using the results of \cite{massSweden}, which hugely advanced the former, this paper
takes the first steps towards accomplishing the latter. More specifically, using the
Berends-Giele-like construction of the $n$-point tree-level string amplitudes involving one
massive and $n{-}1$ massless states \cite{massSweden}, we identify a precise relation with the $\ap^2$ sector
of the $(n{+}1)$-point massless tree-level amplitudes.

The underpinnings of this relation rely on
the combinatorially-rich objects dubbed {\it scalar BRST invariants}. They played a major role
in the joint analysis of the $\ap^2$ sector of the massless string tree amplitudes and the
low-energy limit of one-loop open string amplitudes \cite{oneloopbb}.
They are naturally generated using the zero-mode saturation rules in
the pure spinor formalism \cite{psf}
and obey several identities
\cite{partIcohomology}. In addition, they are mysteriously connected to a combinatorial algorithm
\cite{EOMBBs} related to
Bern-Carrasco-Johansson (BCJ)
amplitude relations \cite{BCJ} and appear in the context of the descent algebra \cite{cdescent}.
As we will see below, yet another relation will be added to this list.

Note that the factorization of the massless amplitudes on the massive poles implies that
massive and massless amplitudes are related, see for example \cite{Bianchi:2015yta}.
However, the factorization condition necessarily involves a quadratic expression of massive
amplitudes and, to our knowledge, has never been used to express a single massive amplitude
in terms of massless data.

To avoid index positioning gymnastics, particle labels will be written mostly downstairs and
vector indices mostly upstairs. Repeated indices are summed over and $[m_1
\ldots m_N]$ does not contain $1/N!$.

\section{String scattering with massive states}

The bosonic physical states at the first massive level of the superstring are described
by a symmetric traceless tensor $g_{mn}$ and a $3$-form $b_{mnp}$ 
of SO(10) subject to
$\p^m g_{mn}=\p^m b_{mnp}=0$ and comprising $44$ and $84$ degrees of freedom, respectively.

In a recent paper \cite{massSweden}, the superstring amplitude involving $n{-}1$ massless states
and one massive state $\underline{n}$ was packaged in terms of $(n{-}3)!$
worldsheet integrals $F^P_Q$ and partial subamplitudes $A(1,P,n{-}1|\underline{n})$ as
\beq
\label{stringmass}
{\cal A}(1,Q,n{-}1,\underline{n}) = \sum_{P \in S_{n-3}} F^P_Q
A(1,P,n{-}1|\underline{n})\,,
\eeq
where $P$ and $Q$ are words comprised of particle labels (letters) and $F^P_Q$
have the same functional form as the string disk integrals
in the massless string scattering amplitude \cite{MSSI,MSSII,drinfeld,BGap};
the only difference stems from the massive constraint $k_{\underline{n}}^2 =
-1/\ap$ affecting the relations among Mandelstam variables.
These integrals will play no role in the discussions below, and
we will focus
our attention in the partial amplitudes $A(1,P|\underline{n})$.

When all external states are bosonic,
the partial amplitude with $|P|=n{-}1$ massless states and one state from the
first massive multiplet
are given by \cite{massSweden}
\beq\label{APun}
A(P|{\underline n}) = \phi_P^{mn} g_{\underline n}^{mn}
+ \phi_P^{mnp}b_{\underline n}^{mnp}\,,
\eeq
where $g_{mn}$ and $b_{mnp}$ are the massive polarizations while
the $n{-}1$ massless states are encoded in (note the modified
normalization conventions compared to \cite{massSweden}):
\begin{align}
\label{allmul}
\phi_P^{mn} &= \ap\!\!\!\! \sum_{XY=P} f_X^{ma}f_Y^{an} + {\rm cyc}(P)\,, \\
\phi_P^{mnp} &= 2i\!\!\!\sum_{XY=P} e_X^m k_Y^n e_Y^p - {4i\over3}\!\!\!
\sum_{XYZ=P} e_X^m e_Y^n e_Z^p + {\rm cyc}(P)\,.\notag
\end{align}
The notation ${}{+}{\rm cyc}(P)$ instructs to add the cyclic permutations
of the letters in $P$, $XY{=}P$ denote the deconcatenations of
$P$ into non-empty words $X$ and $Y$, and $k_{ij \ldots p}^m = k^m_i + k^m_j + \cdots + k^m_p$.
The multiparticle polarizations in \eqref{allmul} obey the recursion \cite{BGBCJ} (equivalent
to \cite{BG})
\begin{align}
\label{multSYM}
e_P^m &= {1\over k_P^2}\sum_{XY=P}\Big[ e^m_Y(k_Y\cdot e_X) + f_X^{mn}e_Y^n -
(X\leftrightarrow Y)\Big]\,,\notag\\
f_P^{mn} &= k_P^m e_P^n - k_P^n e_P^m -\!\!\! \sum_{XY=P}\big(e_X^m e_Y^n - e_X^n e_Y^m\big)\,,
\end{align}
starting with the single-particle $e_i^m$ gluon polarization vector
and its field strength $f_i^{mn}=k^m_i e_i^n - k^n_i e_i^m$.

Example amplitudes for $\underline{n}=3,4$ read
\begin{align}
\label{ampsM}
A(1,2|\underline{3}) &=
       i e_{1}^m f_2^{np}b_{\underline{3}}^{mnp}
       + \ap f_1^{mp} f_2^{pn} g_{\underline{3}}^{mn}
       + {\rm cyc}(12)\notag\\
A(1,2,3|\underline{4}) &=
i\big(
       2 e_{1}^m k_{23}^n e_{23}^p
       + 2 e_{12}^m k_3^n e_{3}^p
       - {4\over 3} e_{1}^m e_{2}^n e_{3}^p
       \big)b_{\underline{4}}^{mnp}\notag\\
       &\!+ \ap \big(f_1^{ma} f_{23}^{an}
       + f_{12}^{ma} f_3^{an}\big)g_{\underline{4}}^{mn}
       + {\rm cyc}(123)
\end{align}

\medskip\noindent
{\bf Massless strings at $\ap^2$ order:}
Recall the definition of $A^{F^4}$ as
the {\it massless} disk amplitudes at $\ap^2$ order
\cite{oneloopbb}
\beq\label{AFqdef}
A(Q) = A^{\rm YM}(Q) + \ap^2 \zeta_2 A^{F^4}(Q) + \cdots\,.
\eeq
We will now propose a map that
replaces the massive external state $\underline{n}$ by two {\it massless} states $n$ and $n{+}1$
whose momenta satisfy $2\ap(k_n\cdot k_{n{+}1}){=-}1$. It turns
the massive $n$-point amplitude $A(P|\underline{n})$ into sums of {\it massless\/}
$\ap^2 A^{F^4}$ at $n{+}1$
points. For convenience, let us use the shorthand $H$ for this map.
More precisely,
\beq\label{mapH}
H:
\begin{cases}
(g_{\underline n}^{rs},b_{\underline n}^{rst})&\rightarrow
(g^{rs}_{n,n{+}1}, b^{rst}_{n,n{+}1})\,,\\
\ap k_{\underline{n}}^2 =-1&\rightarrow 2\ap(k_n\cdot k_{n{+}1}) = -1
\end{cases}
\eeq
with
\begin{align}
\label{masspols}
g^{rs}_{n,n{+}1} &= {1\over8}\big( e_n^r e_{n{+}1}^s
          + e_n^s e_{n{+}1}^r
          - {1\over3} \d^{rs} (e_n\cdot e_{n{+}1})\big)\\
& + {\ap\over12}  \Big(
          (k_n^r k_n^s
          - 2 k_n^r k_{n{+}1}^s) (e_n\cdot e_{n{+}1})\cr
&\qquad\quad{} + 3 (k_{n{+}1}^r e_n^s + k_{n{+}1}^s e_n^r) (k_n\cdot e_{n{+}1})\cr
&\qquad\quad{}+ (n\leftrightarrow {n{+}1}) \Big)\cr
& -{\ap\over12} \d^{rs} (k_n\cdot e_{n{+}1}) (k_{n{+}1}\cdot e_n)\cr
&       + {\ap^2\over6} k_{n,n{+}1}^r k_{n,n{+}1}^s(k_n\cdot e_{n{+}1})(k_{n{+}1}\cdot e_n)\,,\cr
b^{rst}_{n,{n{+}1}} &=
       {i\ap\over16}  \big(
       k_n^{[r}e_n^s e_{n{+}1}^{t]} + k_{n{+}1}^{[r}e_{n{+}1}^s e_n^{t]}
          \big)\cr
       &+ {i\ap^2\over8} \big(
        k_n^{[r}k_{n{+}1}^s e_{n{+}1}^{t]}(k_{n{+}1}\cdot e_n)\cr
&\qquad\quad{} + k_{n{+}1}^{[r}k_n^s e_n^{t]}(k_n\cdot e_{n{+}1})
          \big)\,,\notag
\end{align}
For example, with $s_{ij}=(k_i\cdot k_j)$
\begin{align}
\label{examp}
A(1,2|\underline{3})\Big|_H &= - \ap^2 A^{F^4}(1,2,3,4),\;\; s_{34}=-{1\over 2\ap}\\
A(1,2,3|\underline{4})\Big|_H &=
\ap^2 \AFq(1,3,4,2,5)
        - \ap^2\AFq(1,4,2,3,5)\notag\\
        &- \ap^2\AFq(1,2,5,3,4)\,,\;\; s_{45}=-{1\over 2\ap}\,.\notag
\end{align}
In general,
\beq\label{CtoAFq}
A(1,P|{\underline n})\Big|_H = -{\ap^2\over6}A^{F^4}(\g_{1|P,n,n{+}1}),\;\; s_{n,n{+}1}=-{1\over 2\ap}
\eeq
where $\g_{1|P,n,n{+}1}$ are the
{\it BRST-invariant permutations} related to the descent algebra defined
in \cite{cdescent}.\footnote{Note that \eqref{CtoAFq} is not written in a minimal basis of
$A^{F^4}$ amplitudes. Additional KK-like relations \cite{oneloopbb,cdescent}
were used to arrive at the examples \eqref{examp}.}
We have explicitly \cite{FORM} checked the validity of \eqref{CtoAFq} up to $\underline{n}=6$.


The consistency of \eqref{masspols} can be verified from
$k_{ij}^m g^{mn}_{ij}=k_{ij}^m b^{mnp}_{ij}=0$, and that $g_{mn}$
is traceless symmetric while $b_{mnp}$ is totally antisymmetric.
To see this, one uses
the transversality $(k_i\cdot e_i)=0$ and the mass $k_i^2{=}k_j^2{=}0$
of the gluon states
and the constraint $2\ap (k_i\cdot k_j) = -1$.

\section{Derivation}

The derivation of the relations \eqref{masspols} and \eqref{CtoAFq} are the result
of an alternative construction of a superstring massive
vertex operator and its subsequent use in an amplitude calculation at tree level using the pure
spinor formalism. In the following discussions we will briefly outline the techniques and reasoning
that led to those relations. More
details will appear in a longer paper \cite{longer}.

\medskip\noindent
{\bf CFT basics of the pure spinor formalism:}
The pure spinor formalism \cite{psf} is based on a conformal field theory (CFT) on the two-dimensional
string worldsheet. As such, the prescription to compute tree-level amplitudes of string states
is given by a correlation function of vertex operators inserted at points $z_i$ on
a genus-zero Riemann surface
\beq\label{amppresc}
A = \langle V_1(z_1)V_2(z_2)V_3(z_3) \prod_{i=4}^n\int dz_i U_i(z_i)\rangle
\eeq
where the brackets $\langle - \rangle$ indicate a CFT correlation function (see \cite{treereview}
for a review).
The integrated (unintegrated) vertices $\int U_i$ ($V_i$) for
physical states at the mass level $n$
are ghost-number zero (one) expressions
in the cohomology of the pure spinor BRST charge, $Q=\oint \l^\a d_\a$,
with conformal weight $n{+}1$ ($n$) at zero momentum. $\l^\a$ is a bosonic spinor satisfying the
pure spinor constraint $(\l\g^m\l)=0$ and $d_\a$ is the supersymmetric Green-Schwarz constraint.
Finally, after integrating out the variables of non-vanishing
conformal weight (see below), the amplitude prescription \eqref{amppresc} reduces to a correlation involving
only the zero-modes of
$\l^\a$ and $\t^\a$.
They are integrated out using the prescription
$\langle (\l\g^m\t)(\l\g^n\t)(\l\g^p\t)(\t\g_{mnp}\t)\rangle =
2880\ap^2$.

\medskip\noindent{\bf Massless vertices:}
The vertex operators for the massless states are given by \cite{psf}
\begin{align}\label{vertices}
V&=\lambda^\alpha A_\alpha\,,\\
U&=\p\theta^\alpha A_\alpha+\Pi^mA_m
+2\alpha'd_\alpha W^\alpha+\alpha'N^{mn}F_{mn}\,,\notag
\end{align}
where $A_\a$, $A^m$, $W^\a$ and $F^{mn}$ are the
ten-dimensional\footnote{Recall that ten-dimensional superspace is described
by $X^m$ with $m=1, \ldots,10$ and $\t^\a$ with $\a=1, \ldots,16$.} linearized superfields
describing the SYM multiplet while $\Pi^m$ is a supersymmetric momentum and $N^{mn}$ is the
Lorentz current of the pure spinor. The superfields satisfy
\cite{wittentwistor}
\begin{align}
\label{symeom}
&D_{(\a} A_{\b)} =\gamma^m_{\alpha\beta}A_m,\quad
D_\alpha A_m = (\gamma^m W)_\a + \p_m A_\a,\\
&D_\a W^\b = \frac{1}{4}(\g^{mn})_\a{}^\b F_{mn},\quad
D_\a F_{mn} = \p_{[m}(\g_{n]} W)_\a\,.\notag
\end{align}
The variables $\l^\a$, $\t^\a$ ($\p\t^\a$, $\Pi^m$, $d_\a$ and $N^{mn}$)
have conformal weight zero (one). Thus,
the massless vertices \eqref{vertices} have conformal weights zero and one, respectively. Furthermore,
the equations of motion \eqref{symeom} imply $QV=0$ and $QU=\p V$.

\medskip\noindent
{\bf Massive unintegrated vertex:}
The unintegrated vertex operator for the first massive level
was constructed using ten-dimensional superspace in \cite{BC},
\begin{gather}
\label{unmass}
V=\l^\a\p\t^\b B_{\a\b} + \l^\a\Pi^m H_{\a m} + 2\ap \l^\a d_\b C^\b{}_\a\\
+ \ap N^{mn}\l^\a F_{\a mn}\,,\notag
\end{gather}
where $B_{\a\b}$, $H^m_\a$, $C^\b{}_\a$ and $F_{\a mn}$ are superfields encoding the massive
polarization tensors (and spinors) of the first massive supermultiplet.
Their equations of motion were spelled out in \cite{BC} and they were gauge fixed to
\begin{gather}
\label{BCgauge}
B_{\a\b} = \g^{mnp}_{\a\b}B_{mnp}\,,\quad  \p^m B_{mnp} = 0\,,\\
\g^{m\a\b}H_{m\b} = 0\,,\quad \p^m H_{m\a} = 0,\notag\\
C^\a{}_\b =
{1\over4}(\g^{mpnq})^\a{}_\b\p_{m}B_{npq}\,,\quad
\g^{m\a\b}F_{\a mn} = 0\,.\notag
\end{gather}
The construction in \cite{BC} followed a general ansatz with the correct conformal weight
and ghost number and BRST invariance $QV=0$. Alternatively, one can
derive a mass-level $n$ unintegrated vertex $V$
using the OPEs between the massless vertices. The
prescription is \cite{FMS,Chakrabarti}
\beq\label{Vmass}
V_3(z) =\oint_z dw\, U_1(w)V_2(z)\,,\quad 2\ap(k_1\cdot k_2) = -n,
\eeq
where $U_1$ and $V_2$ are integrated and unintegrated {\it massless} vertices
containing the plane waves $e^{ik_1\cdot X}$ and $e^{ik_2\cdot X}$ with $k_1^2{=}k_2^2{=}0$.
Under the OPE, the plane waves of $U_1$ and $V_2$ combine to the plane wave $e^{i(k_1+k_2)\cdot X}$ of $V_3$,
and the n$^{\rm th}$ mass-level condition $(k_1+k_2)^2 = -n/\ap$ gives rise to the constraint
$2\ap(k_1\cdot k_2) = -n$, ensuring that the contour integral picks up the correct conformal
weight.
It follows from \eqref{Vmass} with $n{=}1$ that
$V_3(z)$ is BRST invariant, has ghost number one, and has conformal weight one at zero
momentum. Therefore, it qualifies to be
an unintegrated vertex operator for the first massive level.

Long but straightforward calculations using the OPEs between massless vertices yield an expression
for the massive vertex \eqref{Vmass} with the following massless SYM representation for the massive
superfields:
\begin{align}
\label{symreps}
B_{\a\b}&= -2\ap i k_2^m(\g^m W_1)_\b A^2_\a - \ap i k_1^m (\g^n W_1)_\b
(\g^{mn}A_2)_\a\cr
&- {\ap\over2}F_1^{mn}(\g_{mn}D)_\b A^2_\a\,, \\
H^m_\a &=
A_{1}^m A^{2}_\a
	+ 2\ap k_1^m(k^2\cdot A^{1}) A^{2}_\a \cr
&-2 i\ap  k_1^m W^\b_1D_\b A^{2}_\a
	- {\ap\over2}ik_1^m F_{1}^{np}(\gamma_{np}A_2)_\a\,, \cr
C^\b{}_\a &=W_1^\b A^{2}_\a\,, \cr
F_{\a mn} &=F^{1}_{mn}A^{2}_\a\,.\notag
\end{align}

\medskip\noindent{\bf Gauge fixing:} While the vertex operator \eqref{Vmass}
with the explicit SYM realization \eqref{symreps} of its superfields is a legitimate unintegrated
vertex operator, it still contains gauge redundancies due to $V_3\rightarrow V_3+Q\Omega$ that
need to be fixed. Following the gauge-fixing procedures of \cite{BC},
a long set of redefinitions detailed in \cite{longer} yields the massless SYM
representation of the massive superfields satisfying the gauge conditions \eqref{BCgauge}:
\begin{align}
\label{symrepsBC}
B_{mnp} &= \frac{1}{18}\ap (W_1\g_{mnp}W_2)
+\frac{1}{9}\ap ^2k^{1}_{[m}k^{2}_{n}(W_{1}\g_{p]}W_{2})\cr
& +\frac{1}{18}i\ap^2\Big[k^{2q}F^{1}_{q[m}F^{2}_{np]} +(1\leftrightarrow 2) \Big]\,,\\
H_{m\a} &=
	 {i\ap\over6}\Bigl(-5i  F^{1}_{mn}(\g^nW_{2})_\a
	-2 k^{12}_mA^{1}_n(\g^n W_2)_\a \cr
&\hspace{1cm} +  k^1_pA^{1}_n(\g^{mnp}W_{2})_\a\cr
&\hspace{1cm}
	-4\ap k^{12}_m(k^2\cdot A^{1})k^1_n(\g^nW_{2})_\a \
	+\ (1\leftrightarrow 2) \Bigr)\,,\cr
C^\b{}_\a &={1\over4}(\g_{mnpq})^\b{}_\a ik_{12}^m B^{npq}\,,\cr
F_{\a mn} &= \frac{1}{16}\Big(
7ik^{12}_{[m}H_{n]\a}
 + ik^{12}_q(\g_{q[m})_\a{}^\b H_{n]\b}\Big)\,,\notag
\end{align}
with $B_{\a\b}=\g^{mnp}_{\a\b}B_{mnp}$.

\medskip\noindent{\bf The massive polarization map
\eqref{masspols}:} We are now in a position to explain the origin
of the prescription \eqref{masspols}. According to the $\t$ expansion analysis of the
massive superfields \cite{masstheta},
the massive polarizations $g_{mn}$ and $b_{mnp}$ can be extracted
from the massive superfields as
\beq\label{polpresc}
g^{mn} = {1\over64}(D\g^{(m}H^{n)})\big|_{\t=0},\quad
b^{mnp} = {9\over8}B^{mnp}\big|_{\t=0}\,,
\eeq
where the overall normalizations were chosen for later convenience. The
expressions in \eqref{masspols} follow from the above definitions using the
massless representations \eqref{symrepsBC}.

The origin of \eqref{CtoAFq} will become clear in the following discussion of
the three-point amplitude.

\medskip\noindent{\bf Three-point tree amplitude:}
The string three-point amplitude with one massive and two massless states was firstly computed in
the pure spinor formalism in \cite{PSthreemass} and simplified in
\cite{unnotes}:
\beq\label{3ptmass}
A(1,2|{\underline 3}) ={i\over2\ap} \langle V_1 (\l\g_m W_2)(\l H^m_3)\rangle\,,
\eeq
where particles $1$ and $2$ are massless SYM states and $3$ is massive. The component expansion in
terms of polarization and momenta
of \eqref{3ptmass} can be evaluated in two different ways:
\begin{enumerate}
\item Using the theta expansion of the
massive superfield $H^m_{3\,\a}$ in terms of $g_{mn}$ and $b_{mnp}$ derived in
\cite{masstheta}. This yields the expression in \eqref{ampsM}.

\item Using the massless SYM representation of $H^m_{3\,\a}$ and performing the
calculations as a regular four-point pure spinor superspace expression, while
imposing the constraint $2\ap(k_3\cdot k_4) = -1$ after the last step.
This yields (with $s_{ij}=(k_i\cdot k_j)$),
\end{enumerate}
\begin{gather}
{1\over \ap^2}A(1,2|\underline{3}) = \\
s_{23} \Big(
           (k_1\cdot e_2) (k_1\cdot e_3) (e_1\cdot e_4)
          - (k_1\cdot e_2) (k_1\cdot e_4) (e_1\cdot e_3)\cr
          + (k_1\cdot e_2) (k_2\cdot e_3) (e_1\cdot e_4)
          - (k_1\cdot e_2) (k_2\cdot e_4) (e_1\cdot e_3)\cr
          - (k_1\cdot e_2) (k_3\cdot e_1) (e_3\cdot e_4)
          - (k_1\cdot e_3) (k_2\cdot e_1) (e_2\cdot e_4)\cr
          + (k_1\cdot e_3) (k_2\cdot e_4) (e_1\cdot e_2)
          + (k_1\cdot e_4) (k_2\cdot e_1) (e_2\cdot e_3)\cr
          - (k_1\cdot e_4) (k_2\cdot e_3) (e_1\cdot e_2)
          - (k_2\cdot e_1) (k_2\cdot e_3) (e_2\cdot e_4)\cr
          + (k_2\cdot e_1) (k_2\cdot e_4) (e_2\cdot e_3)
          + (k_2\cdot e_1) (k_3\cdot e_2) (e_3\cdot e_4)\cr
	  - (e_1\cdot e_2) (e_3\cdot e_4) s_{23}
          \Big)\cr
	  + s_{12}  \Big(
          (k_1\cdot e_2) (k_2\cdot e_3) (e_1\cdot e_4)
          - (k_1\cdot e_3) (k_3\cdot e_2) (e_1\cdot e_4)\cr
          + (k_1\cdot e_4) (k_2\cdot e_1) (e_2\cdot e_3)
          - (k_1\cdot e_4) (k_2\cdot e_3) (e_1\cdot e_2)\cr
          + (k_1\cdot e_4) (k_3\cdot e_2) (e_1\cdot e_3)
          - (k_2\cdot e_1) (k_2\cdot e_3) (e_2\cdot e_4)\cr
          + (k_2\cdot e_1) (k_2\cdot e_4) (e_2\cdot e_3)
          + (k_2\cdot e_1) (k_3\cdot e_2) (e_3\cdot e_4)\cr
          - (k_2\cdot e_3) (k_3\cdot e_1) (e_2\cdot e_4)
          + (k_2\cdot e_4) (k_3\cdot e_1) (e_2\cdot e_3)\cr
          + (k_3\cdot e_1) (k_3\cdot e_2) (e_3\cdot e_4)
	  - (e_1\cdot e_4) (e_2\cdot e_3) s_{12}
          \Big)\cr
	  + s_{12} s_{23}   \Big(
	  (e_1\cdot e_3) (e_2\cdot e_4)
          - (e_1\cdot e_2) (e_3\cdot e_4)\cr
          - (e_1\cdot e_4) (e_2\cdot e_3)
          \Big)\notag
\end{gather}
which, before imposing the constraint $2\ap s_{12}{= -}1$, is
readily recognized as $-A^{F^4}(1,2,3,4)$, the $\ap^2$ correction to the massless
four-point string amplitude \eqref{AFqdef}.

Since both ways compute the same amplitude, there must be a
correspondence between them. Looking for a similar pattern at higher points
led to the proposal \eqref{CtoAFq}.

It turns out that capturing the general pattern is easier using
the scalar BRST invariants defined in \cite{oneloopbb,EOMBBs} and whose bosonic components are available
to download from
\cite{website}. Starting from \eqref{APun} and using the map \eqref{mapH},
we explicitly checked that:
\begin{align}
\label{AtoCs}
A(1,2|{\underline3})\Big|_H &= -\ap^2\langle C_{1|2,3,4}\rangle,\;\; s_{34}=-{1\over 2\ap}\\
A(1,2,3|{\underline4})\Big|_H &= -\ap^2\langle C_{1|23,4,5}\rangle,\;\; s_{45}=-{1\over 2\ap}\cr
A(1,2,3,4|{\underline5})\Big|_H &= -\ap^2\langle C_{1|234,5,6}\rangle,\;\; s_{56}=-{1\over 2\ap}\cr
A(1,2,3,4,5|{\underline6})\Big|_H &= -\ap^2\langle C_{1|2345,6,7}\rangle,\;\; s_{67}=-{1\over 2\ap}\notag
\end{align}
which, in turn, suggests the generalization
\beq\label{relmm}
A(1,P|{\underline n})\Big|_H = -\ap^2\langle C_{1|P,n,n{+}1}\rangle\,,\;\;
s_{n,n{+}1}=-{1\over2\ap}\,.
\eeq
The translation to linear combinations of $A^{F^4}$
amplitudes in \eqref{CtoAFq} follows from the permutations $\g$ of
\cite{cdescent}
\beq
\langle C_{1|P,Q,R}\rangle = {1\over 6}A^{F^4}(\g_{1|P,Q,R}).
\eeq
Further evidence for \eqref{relmm} stems from the fact that both sides are
annihilated by shuffling $P=R\shuffle S$ for non-empty $R$ and $S$; the left-hand side due to the
Kleiss-Kuijf identity of the massive partial amplitude \cite{massSweden}, and the right-hand side
by construction \cite{oneloopbb,EOMBBs}. Therefore, we uncovered a hidden relation between the massive
string tree amplitude with one massive external state and the $\ap^2$ sector
of the purely {\it massless} tree-level string amplitudes.

\section{\label{sec:outline} Conclusion and outlook}

In this paper we found a relation between the $n$-point string tree amplitude with one massive and
$n{-}1$ massless states and linear combinations  of $n{+}1$ massless string tree amplitudes at
$\ap^2$ order. To see this, we defined a map that replaces the massive polarizations of one massive
leg by the polarizations and momenta of two massless gluons. Then, after being transformed by this map,
the partial amplitudes $A(P|\underline{n})$ of
the full string tree amplitude \eqref{stringmass},
are written in terms of the $\ap^2$ correction of the purely
massless string disk amplitude.

It is not the first time that relations were discovered where
some string states are replaced by others: the prime example being the KLT relations at
tree level trading one graviton for two gluons \cite{KLT}, see also
\cite{Stieberger:2016lng,Stieberger:2021daa,Mazloumi:2022lga,Stieberger:2022lss} for relations along the same lines.
However, the relation found in this paper not only trades massive for massless polarizations
but also connects amplitudes at different orders of $\ap$ expansions.

It will be interesting to extend the observations here to more external massive states as they
will probably give rise to linear combinations of amplitudes at higher $\ap$ orders. How to
characterize the associated permutations? Another question
to investigate is related to the factorization on massive poles of the massless tree amplitudes
\cite{Bianchi:2015yta}. Using the results presented here
could lead to some sort of
self consistency built in in the massless tree amplitudes via their $\ap$
expansion.
Moreover, similar relations are also expected to hold in bosonic string amplitudes, where a wealth of data
is available \cite{Huang:2016tag,Azevedo:2018dgo}.

Also, it is worth noting that there are more ``topologies'' of the scalar BRST invariants starting
at multiplicity six; for example, $C_{1|234,5,6}$ and $C_{1|23,45,6}$. They have different
combinatorial properties and their expansions in terms of $A^{F^4}$ are completely different. As
already explicitly checked in \eqref{AtoCs}, only one topology appears at (massless) multiplicities six and
seven. In general, what
happens to the other topologies? Do they map to something meaningful?

Finally, it would be desirable to invert the map \eqref{mapH} as a means of
obtaining the massive string amplitudes starting from their massless counterparts.
If this is achieved and extensions with more massive legs and higher orders in $\ap$ are found,
it would mean that all massive amplitudes could be simply extracted from the massless amplitudes
computed in
\cite{MSSI,MSSII,drinfeld,BGap}.

\medskip
\noindent\textit{Acknowledgements:} We thank
Oliver Schlotterer, Paolo Di Vecchia, and Nathan Berkovits for comments on the draft, and
Bruno Rodrigues Soares for discussions.
SPK would
like to thank IMSc for its support and hospitality, where parts of the work
were done. CRM is supported by a University Research Fellowship from the Royal Society.
MV was also supported in part by the STFC consolidated grant ST/T000775/1
``New Frontiers in Particle Physics, Cosmology and Gravity''.
During the initial stages of this work LAY was supported by CRM's Royal Society University
Research Fellowship.

\end{document}